\documentclass[oribibl]{llncs}
\usepackage{multicol}
\usepackage{amsmath}
\usepackage{amssymb}
\usepackage{graphicx}
\usepackage{epsfig}
\usepackage{color}
\usepackage{hhline}

\usepackage{hyperref}

\usepackage{xspace,epic,eepic}

\newcommand{\ignore}[1]{}

\newcommand{\mbf}[1]{\mathbf{ #1}}
\newcommand{\e}{\mathbf{e}}

\def\Li{\mathrm{\Phi}}

\newcommand{\boxtheorem}{  \hfill $\Box$\\}
\newcommand{\nit}[1]{{\it #1}}

\newcommand{\mc}[1]{\mathcal{ #1}}

\newcommand{\shap}{{\sf SHAP}}
\newcommand{\shp}{\#\text{{\sc P}}}
\newcommand{\es}{\mathbf{e}}

\title{Score-Based Explanations in Data Management and Machine Learning}

\author{{\bf Leopoldo Bertossi}\thanks{Email: \ leopoldo.bertossi@uai.cl}}
\institute{\bf Universidad Adolfo Ib\'a\~nez \& Data Observatory Foundation \& IMFD\\
\ignore{Faculty of Engineering and Sciences,} Santiago, \ Chile}

\begin{document}

\thispagestyle{empty}
\pagestyle{plain}
\maketitle

\begin{abstract}
We describe some approaches to  explanations for observed outcomes in data management and machine learning. They are based on the assignment of numerical scores to predefined and potentially relevant inputs. More specifically, we consider explanations for query answers in databases, and for results from classification models. The described approaches are mostly of a causal and counterfactual nature. We argue for the need to bring domain and semantic knowledge into score computations; and suggest some ways to do this.
\end{abstract}

\section{Introduction}

In data management and machine learning one wants {\em explanations} for certain results. For example, for query results from databases, and for outcomes of classification models. Explanations, that may come in different forms, have been the subject of philosophical enquires for a long time, but, closer to home, they appear under different forms in model-based diagnosis and in causality as developed in artificial intelligence. In the last few years, explanations that are based on {\em numerical scores} assigned to elements of a model that may contribute to an outcome have become popular. These scores attempt to capture the degree of their contribution to an outcome, e.g. answering questions like these: What is the contribution of this tuple to the answer to this query? \ What is the contribution of this feature value of an entity to the displayed classification of the latter?

Let us consider, as an example, a financial institution that uses a
learned classifier, e.g. a decision tree, to determine if clients
should be granted loans or not, returning labels $0$ or $1$, resp. A
particular client, an entity $\es$, applies for a loan, the classifier
returns $M(\es) =1$, i.e. the loan is rejected. The client requests
an explanation. A common approach consists in giving scores to the feature values in $\es$,
to quantify their relevance in relation to the classification
outcome. The higher the score of a feature value, the more explanatory  is that value. For example, the fact that the client has value ``5 years"
for feature {\em Age} could have the highest score.

Motivated, at least to a large extent, by the trend towards {\em explainable AI} \cite{molnar}, different explanation scores have been proposed in the literature. Among them, in data management, the {\em responsibility score} as found in actual causality \cite{HP05,CH04} has been used to
quantify the strength of a tuple  as a cause for a query result \cite{suciu,tocs}.  The Shapley-value, as found in coalition game theory, has been used for the same purpose \cite{LBKS20}.
In machine learning, in relation to results of classification models, the Shapley-value has been used to assign scores to feature values. In the form of the $\shap$-score, it has become quite popular and influential \cite{LetA20,LL17}. A responsibility-based score, ${\sf RESP}$, was introduced in \cite{BLSSV20} to assign numbers to feature values of entities under classification.  It is
based on the notions of counterfactual intervention and causal responsibility.

Some scores used in machine learning
appeal to the components of the mathematical model behind the
classifier. There can be all kinds of explicit models, and some are
easier to understand or interpret or use for this purpose. For
example, the FICO-score proposed in \cite{CetA18}, for the FICO dataset
about loan requests, depends on the internal outputs and displayed
coefficients of two nested logistic regression models. Decision trees
\cite{M97}, random forests \cite{BetA84}, rule-based
classifiers, etc., could be seen as relatively easy to understand and use for
explanations on the basis of their components.

Other scores can be applied with {\em black-box} models, in that they
use, in principle, only the input/output relation that represents the
classifier, without having access to the internals details of the
model. In this category we could find classifiers based on complex
neural networks, or XGBoost \cite{LHdR19}. They are opaque enough to
be treated as black-box models. The $\shap$-score and the ${\sf RESP}$-score can be applied to this category.
\ In
\cite{BLSSV20}, the $\shap$-score, the ${\sf RESP}$-score and the FICO-score
are compared. \ In general, the computation of  the first two is intractable.

The $\shap$-score and the ${\sf RESP}$-score can be applied with open-box models.  In this case, an interesting  question is whether having access to the
mathematical model may make their computation tractable, at least for some classes of classifiers.

As suggested above, scores can be assigned to tuples in databases, to measure their contribution to a query answer, or to the violation of an integrity constraint. The responsibility score has been applied for this purpose \cite{suciu,tocs}, and is based on causality in databases \cite{suciu}. Also the Shapley-value has been used for this task \cite{LBKS20}.

In this article we survey some of the approaches to score-based explanations we just mentioned above, in databases and in classification in machine learning. This is not intended to be an exhaustive survey of these areas, but it is heavily influenced by our latest research. Next, we discuss the relevance of bringing {\em domain and semantic knowledge} into these score computations. We also show some first ideas and techniques on how this knowledge can be accommodated in the picture.
 To introduce the concepts and techniques we will use mostly examples, trying  to convey the main intuitions and issues.

This paper is structured as follows. In Section \ref{sec:dbs} we concentrate on causal explanations in databases. In Section \ref{sec:shapy}, we describe the use of the Shapley-value to provide explanation scores in databases.
In Section \ref{sec:shapy}, we describe score-based explanations for classification results. In Section \ref{sec:cons}, we show how semantic knowledge can be brought into the score computations. We conclude with some final remarks in Section \ref{sec:last}.

\section{Explanations in Databases}\label{sec:dbs}

 In data management we {need to understand and compute}
{\em  why}  certain results are obtained or not, e.g. query answers,  violations of semantic conditions, etc.; and we
expect a  database system to provide {\em explanations}.

\subsection{Causal responsibility}\label{sec:causal}

Here, we will consider
{\em causality-based explanations}  \cite{suciu,suciuDEBull}, which we will illustrate by means of an example.

 \begin{example}  \label{ex:uno}  \ Consider the database ${D}$, and the Boolean conjunctive query (BCQ) 
 \label{`first'}

\begin{multicols}{2}

\hspace*{1cm}{\small {\begin{tabular}{l|c|c|} \hline
$R$  & $A$ & $B$ \\\hline
 & $a$ & ${b}$\\
& $c$ & $d$\\
& ${b}$ & ${b}$\\
 \hhline{~--}
\end{tabular} \hspace*{0.5cm}\begin{tabular}{l|c|c|}\hline
$S$  & $A$  \\\hline
 & $a$ \\
& $c$ \\
& ${b}$ \\ \hhline{~-}
\end{tabular} } }

  ${\mc{Q}\!: \ \exists x \exists y ( S(x) \land R(x, y) \land S(y))}$.

 \noindent It holds: \ ${D \models \mc{Q}}$, i.e. the query is true in $D$.
\end{multicols}

 We ask about the  causes for $\mc{Q}$ to be true: \
A tuple ${\tau \in D}$ is
{\em counterfactual cause} for  ${\mc{Q}}$ (being true in $D$) if \ ${D\models \mc{Q}}$ \ and \ ${D\smallsetminus \{\tau\}  \not \models \mc{Q}}$.

In this example,   {$S(b)$ is counterfactual cause for $\mc{Q}$}: \ If ${S(b)}$ is removed from ${D}$,
 ${\mc{Q}}$ is no longer true.

Removing a single tuple may not be enough to invalidate the query. Accordingly, a tuple ${\tau \in D}$ is  an {\em actual cause} for  ${\mc{Q}}$
if there  is a {\em contingency set} \ ${\Gamma \subseteq D}$,  such that \ ${\tau}$ \ is a   counterfactual cause for ${\mc{Q}}$ in ${D\smallsetminus \Gamma}$.

In this example,  ${R(a,b)}$ is an actual cause for ${\mc{Q}}$ with contingency set
${\{ R(b,b)\}}$: \ If ${R(a,b)}$ is removed from ${D}$, ${\mc{Q}}$ is still true, but further removing ${R(b,b)}$ makes ${\mc{Q}}$ false.
\boxtheorem \end{example}

Notice that every counterfactual cause is also an actual cause, with empty contingent set.   Actual but non-counterfactual causes need company to invalidate a query result.
 \ Now we ask  how strong are these tuples as causes? \ For this we appeal to the {\em responsibility} of an actual cause ${\tau}$ for ${\mc{Q}}$ \cite{suciu}, defined by:

\vspace{-6mm} \begin{multicols}{2}
\begin{equation*}{\rho_{\!_D}\!(\tau) \ := \ \frac{1}{|\Gamma| \ + \ 1}},
\end{equation*}

\phantom{ooo}

\noindent with  ${|\Gamma|} = $
size of a smallest contingency set for ${\tau}$, \ and  $0$, otherwise.
\end{multicols}

\begin{example} \ (ex. \ref{ex:uno} cont.) \ The {responsibility of ${R(a,b)}$ is \  $\frac{1}{2}$} ${= \frac{1}{1 + 1}}$ \ (its several smallest contingency sets have all size ${1}$).

  ${R(b,b)}$ and ${S(a)}$ are also actual causes with responsibility  \ ${\frac{1}{2}}$; and
  ${S(b)}$ is actual (counterfactual) cause with responsibility \   $1$ ${= \frac{1}{1 + 0}}$. \boxtheorem
\end{example}

High responsibility tuples provide more interesting explanations. Causes in this case are tuples that come with their responsibilities as  ``scores".
Actually, all tuples can be seen as actual causes and only the non-zero scores matter. \ Causality and responsibility in databases can be extended to the attribute-value level \cite{tocs,foiks18}.

 There is a connection between database causality and  {\em repairs} of databases w.r.t. integrity constraints (ICs) \cite{bertossiSynth}, and also  connections to {\em consistency-based diagnosis} \ and \ {\em abductive diagnosis}. These connections have led to new complexity and algorithmic results for causality and responsibility \cite{tocs,flairsExt}. Actually, the latter turns out to be intractable. In \cite{flairsExt}, causality under ICs was introduced and investigated. This allows to bring semantic and  domain knowledge into causality in databases.

 Model-based diagnosis is an older area of knowledge representation where explanations are main characters. In general, the diagnosis analysis is performed on a logic-based model, and certain
elements of the model are identified as explanations. Causality-based explanations are somehow more recent. In this case,
still a model is used, which is, in general, a more complex  than a database with a query.   In the case of databases, actually there is an underlying logical model,  the {\em lineage or provenance} of the query \cite{lineage} that we will illustrate in Section \ref{sec:CE}, but it is still a relatively simple model.

The idea behind {\em actual causality} is the (potential) execution of   {\em counterfactual interventions} on a {\em structural logico-probabilistic model} \ \cite{HP05}, with the purpose of answering hypothetical  or counterfactual questions of the form: \ {\em What would happen if we change ...?}. It turns out that counterfactual interventions can also be used to define different forms of score-based explanations, in the same spirit of causal responsibility in databases (c.f. Section \ref{sec:resp}). \ Score-based explanations can also be defined in the absence of a model, and without counterfactual interventions (or at least with them much less explicit).

\subsection{The causal-effect score}\label{sec:CE}

Sometimes responsibility does not provide intuitive or expected results, which led to the consideration of an alternative score, the {\em causal-effect score}. We show the issues and this score by means of an example.

\begin{example} \ \label{ex:ce} Consider the database ${E}$ that represents the graph below, and the Boolean Datalog query ${\Pi}$ that is true in $E$ if there is a path from ${a}$ to ${b}$. Here, ${E \cup\Pi \models \nit{yes}}$.

\begin{multicols}{3}

 \hspace*{5mm} {\footnotesize \begin{tabular}{l|c|c|} \hline
 {$E$}  &  ${X}$ &  ${Y}$ \\\hline
 {$t_1$} & { $a$} &  {$b$}\\
{$t_2$}&  {$a$} &  {$c$}\\
{$t_3$}&  {$c$} &  {$b$}\\
{$t_4$}&  {$a$} &  {$d$}\\
{$t_5$}&  {$d$} &  {$e$}\\
{$t_6$}&  {$e$} &  {$b$}\\ \cline{2-3}
\end{tabular}}

 \includegraphics[width=3.3cm]{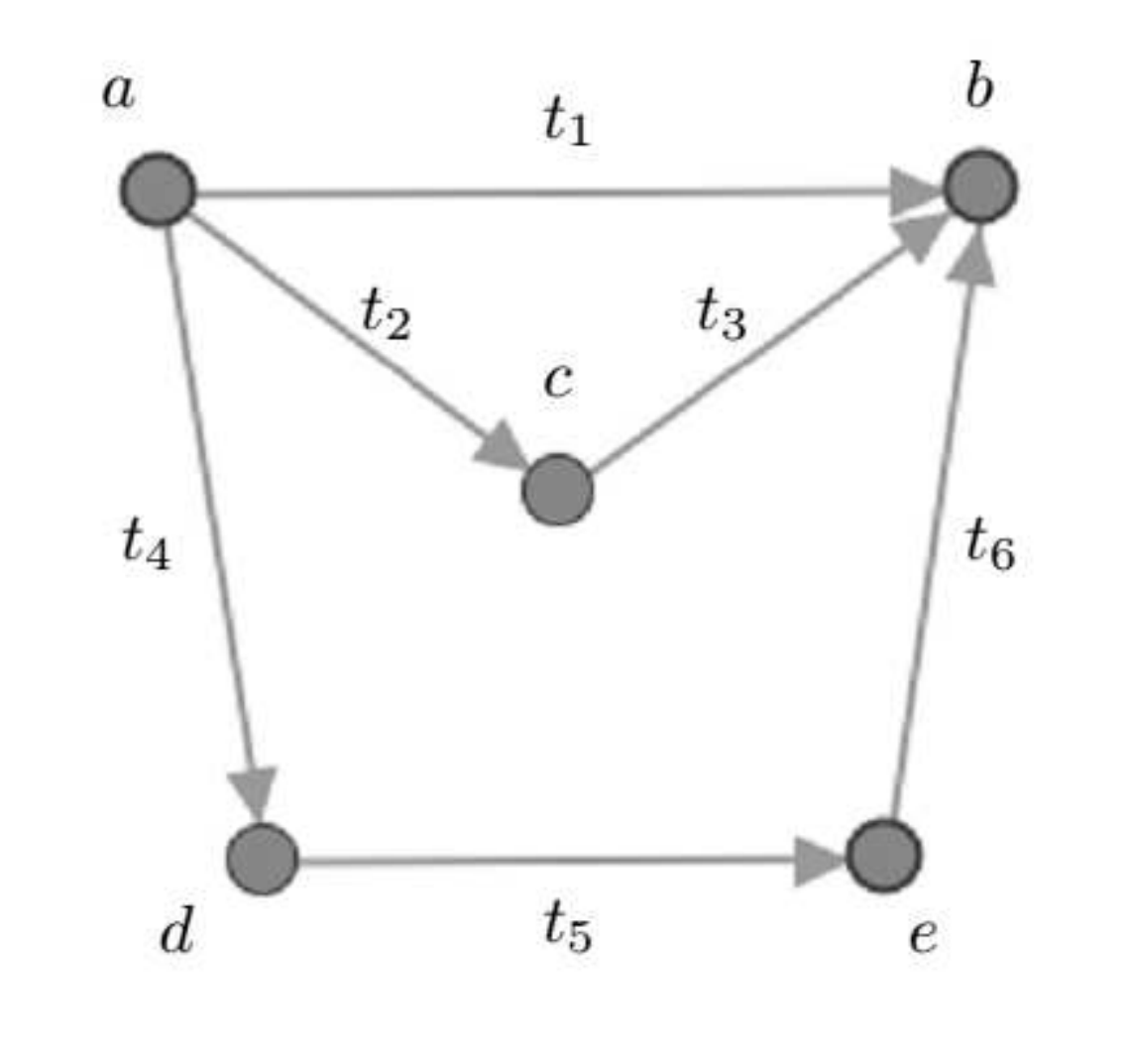}

{\small
\begin{eqnarray}
         \nit{yes }& {\leftarrow}&  {P(a,b)} \nonumber \\
        {P(x,y)}& {\leftarrow}& { E(x,y)} \nonumber \\
        {P(x,y)}& {\leftarrow}& { P(x,z), E(z,y)}\nonumber
\end{eqnarray}
}
\end{multicols}

All tuples are actual causes since every tuple appears in a path from $a$ to $b$. Also,
all the tuples  have the same causal responsibility, {$\frac{1}{3}$}, which
may be counterintuitive, considering that \ ${t_1}$ provides a direct path from ${a}$ to ${b}$.
\boxtheorem\end{example}

In \cite{tapp16}, the notion  {\em causal effect} was introduced. It is based on three main ideas, namely, the transformation, for auxiliary purposes, of the database into a probabilistic database,
interventions on the lineage of the query, and the use of expected values for the query.  This is all shown in the next example.

 \begin{example} \ Consider the database \ ${D}$  below, and a  BCQ. 

\begin{multicols}{2}
{\begin{tabular}{c|c|c|}\hline
 $R$ & $A$ & $B$ \\ \hline
  & $a$ & $b$\\
  & $a$ & $c$\\
  & $c$ & $b$\\ \hhline{~--}
  \end{tabular}~~~~~~~~~\begin{tabular}{c|c|}\hline
 $S$ & $B$ \\ \hline
  & ${b}$\\
  & $c$\\
  & \\ \hhline{~-}
  \end{tabular}}

 \noindent ${\mc{Q}: \ \exists x \exists y (R(x, y) \wedge  S(y))}$, which is true in ${D}$.
\end{multicols}

The lineage of the query { instantiated on ${D}$} is given by the {propositional formula}:
\begin{equation}
{\Li_\mc{Q}(D)= (X_{R(a, b)} \wedge  X_{S(b)})  \vee (X_{R(a, c)} \wedge  X_{S(c)}) \vee (X_{R(c, b)} \wedge  X_{S(b)})}, \label{eq:lin}
\end{equation}
where   ${X_\tau}$ is a  {propositional variable} that is true iff  ${\tau \in D}$. \ Here,
  ${\Li_\mc{Q}(D)}$ \ takes value ${1}$ in ${D}$.

Now, for illustration, we want to quantify the contribution of tuple ${S(b)}$ to the query answer. \
For this purpose, we assign probabilities, uniformly and independently, to the tuples in ${D}$, obtaining a
a {\em probabilistic database} \ ${D^{{p}}}$ \cite{probDBs}.  \ Potential tuples outside ${D}$ get probability $0$.

\vspace{3mm}
   \hspace*{3cm}
   {\footnotesize {
   \begin{tabular}{c|c|c|c|}\hline
 $R^{{p}}$ & $A$ & $B$ & {$\mbox{prob}$}\\ \hline
  & $a$ & $b$ & \scriptsize{${\frac{1}{2}}$}\\
  & $a$ & $c$& \scriptsize{$\frac{1}{2}$}\\
  & $c$ & $b$& \scriptsize{$\frac{1}{2}$}\\ \hhline{~---}
  \end{tabular}~~~~~~~~~\begin{tabular}{c|c|c|}\hline
 $S^p$ & $B$ & $\mbox{prob}$\\ \hline
  & ${b}$& \scriptsize{${\frac{1}{2}}$}\\
  & $c$& \scriptsize{$\frac{1}{2}$}\\
  & & \\ \hhline{~--}
  \end{tabular}}}

 \vspace{2mm}  {The $X_\tau$'s become independent, identically distributed Boolean random variables}; \ and ${\mc{Q}}$ becomes a Boolean random variable.
Accordingly, we can ask about the probability that $\mc{Q}$ takes the truth value $1$ (or $0$) when an {\em intervention} is performed on $D$.

     Interventions are of the form ${\nit{do}(X = x)}$, meaning making ${X}$ take value ${x}$, with $x \in \{0,1\}$, in the {\em structural model}, in this case, the lineage. That is, we ask,
for \ ${\{y,x\} \subseteq \{0,1\}}$, about the conditional probability  ${P(\mc{Q} = y~|~ {\nit{do}(X_\tau = x)})}$, i.e. conditioned to  making ${X_\tau}$ false or true.

For example, with ${\nit{do}(X_{S(b)} = 0)}$ and $\nit{do}(X_{S(b)} = 1)$, the lineage in (\ref{eq:lin}) becomes, resp., and abusing the notation a bit:
\begin{eqnarray*}
\Li_\mc{Q}(D|\nit{do}(X_{S(b)} = 0) &:=&  (X_{R(a, c)} \wedge  X_{S(c)}).\\
\Li_\mc{Q}(D|\nit{do}(X_{S(b)} = 1) &:=& X_{R(a, b)}  \vee (X_{R(a, c)} \wedge  X_{S(c)}) \vee X_{R(c, b)}.
\end{eqnarray*}
On the basis of these lineages and  \ ${D^{{p}}}$, \ when ${X_{S(b)}}$ is made false, \ the probability that the instantiated lineage becomes true in {$D^p$} is:

\vspace{1mm}
  \centerline{${P(\mc{Q} = 1~|~ {\nit{do}(X_{S(b)} = 0)}) = P(X_{R(a, c)}=1) \times P(X_{S(c)}=1) = \frac{1}{4}}$.}
\vspace{1mm}
  Similarly, when ${X_{S(b)}}$ is made true, \ the probability of the lineage  becoming true in ${D^p}$ is:

\vspace{1mm}
  \centerline{${P(\mc{Q} = 1~|~{\nit{do}( X_{S(b)} = 1)}) = P(X_{R(a, b)}  \vee (X_{R(a, c)} \wedge  X_{S(c)}) \vee X_{R(c, b)} =1)}
{=  \  \frac{13}{16}}.$}

\vspace{1mm}
The {\em causal effect} of a tuple ${\tau}$ is defined by:
\begin{equation*}
{\mathcal{CE}^{D,\mc{Q}}(\tau) \ := \ \mathbb{E}(\mc{Q}~|~\nit{do}(X_\tau = 1)) - \mathbb{E}(\mc{Q}~|~\nit{do}(X_\tau = 0))}.
\end{equation*}

In particular, using the probabilities computed so far:
\begin{eqnarray*}
\mathbb{E}(\mc{Q}~|~\nit{do}(X_{S(b)} = 0)) &=&  P(\mc{Q} =1~|~\nit{do}(X_{S(b)} = 0)) \ = \  \frac{1}{4},\\
\mathbb{E}(\mc{Q}~|~\nit{do}(X_{S(b)} = 1)) &=&  P(\mc{Q} =1~|~\nit{do}(X_{S(b)} = 1)) \ = \ \frac{13}{16}.
\end{eqnarray*}

Then, \  the causal effect for the tuple ${S(b)}$ is:
   ${\mathcal{CE}^{D,\mc{Q}}(S(b)) = \frac{13}{16} - \frac{1}{4} = {\frac{9}{16}} \ > \ 0}$, showing that the tuple is relevant for the query result, with a relevance score provided by the causal effect, of   $\frac{9}{16}$.
\boxtheorem \end{example}

Let us now retake the initial example of this section.

\begin{example} \ (ex. \ref{ex:ce} cont.)  \ The Datalog query, here as a union of BCQs, has the lineage: \ ${\Li_\mc{Q}(D) = X_{t_1} \vee (X_{t_2}\wedge X_{t_3}) \vee (X_{t_4} \wedge X_{t_5} \wedge X_{t_6})}.$ It holds:
\begin{eqnarray*}
\mathcal{CE}^{D,\mc{Q}}(t_1)  &=&  {0.65625},\\
\mathcal{CE}^{D,\mc{Q}}(t_2)  &=&  \mathcal{CE}^{D,\mc{Q}}(t_3) =  0.21875,\\
\mathcal{CE}^{D,\mc{Q}}(t_4)  &=&  \mathcal{CE}^{D,\mc{Q}}(t_5) = \mathcal{CE}^{D,\mc{Q}}(t_6) = 0.09375.
\end{eqnarray*}

The causal effects are different for different tuples, and the scores are much more
intuitive than the responsibility scores.  \boxtheorem \end{example}

The definition of the causal-effect score may look  rather {\em ad hoc} and arbitrary. We will revisit it in Section \ref{sec:ester}, where we will have yet another score for applications in databases. Actually, trying to take a new approach to measuring the contribution of a database tuple to a query answer, one can think of applying the {\em Shapley-value}, which  is firmly established in game theory, and also used in several other areas.

The main idea is that {\em several tuples together} are necessary to violate an IC or produce a query result, much like
{players in a coalition game}. Some may contribute more than others to the {\em  wealth distribution function} (or simply,  game function), which in this case becomes the query result, namely $1$ or $0$ if the query is Boolean, or a number if the query is an aggregation.
  The Shapley-value of a tuple can be used to assign a score to its contribution. This was done in \cite{LBKS20}, and will be retaken in Section \ref{sec:ester}. But first things first.

\section{The Shapley-Value in Databases}\label{sec:shapy}

\subsection{The Shapley-Value}
The Shapley value was proposed in game theory by Lloyd Shapley in 1953
\cite{S53}, to quantify the contribution of a player to a coalition game where players share a wealth function.\footnote{The original paper and related ones on the
  Shapley value can be found in the book edited by Alvin Roth
  \cite{R88}. Shapley and Roth shared the Nobel Prize in Economic
  Sciences 2012.} It has been applied in many disciplines. In particular, it has been investigated in computer science under
{\em algorithmic game theory} \cite{DBLP:books/cu/NRTV2007}, and it has been applied to many and
different computational problems. The
computation of the Shapley-value is, in general, intractable. In many
scenarios where it is applied its computation turns out to be
$\shp$-hard \cite{FK92,DBLP:journals/mor/DengP94}.

In particular, the Shapley value has been used in knowledge
representation, to measure the degree of inconsistency of a
propositional knowledge base \cite{HK10}; in data management to
measure the contribution of a tuple to a query answer \cite{LBKS20}  (c.f. Section \ref{sec:ester});
and in machine learning to provide explanations for the outcomes of
classification models on the basis of numerical scores assigned to the participating feature values \cite{LL17} (c.f. Section \ref{sec:shap}).

 Consider a set of players ${D}$,  and a
game function,  $\mc{G}:  \mc{P}(D)  \rightarrow  \mathbb{R}$, where $\mc{P}(D)$ the power set of $D$. \
 The Shapley-value of player ${p}$ in ${D}$ es defined by:
  \begin{equation}{\nit{Shapley}(D,\mc{G},p):= \sum_{S\subseteq
  D \setminus \{p\}} \frac{|S|! (|D|-|S|-1)!}{|D|!}
(\mc{G}(S\cup \{p\})-\mc{G}(S))}.\label{eq:sh}\end{equation}
  Notice that here, ${|S|! (|D|-|S|-1)!}$ is the number of permutations of
${D}$ with all players  in ${S}$  coming first, then ${p}$, and then all the others. That is, this quantity
is the expected contribution of player $p$ under all possible additions of $p$ to a partial random sequence of players followed   by a random sequence of the rests of   the players. Notice the counterfactual flavor, in that there is a comparison between what happens having $p$ vs. not having it. The Shapley-value is the only function that satisfy certain natural properties in relation to games. So, it is a
result of a categorical set of axioms or conditions.

 \subsection{Shapley for query answering}\label{sec:ester}

 Back to query answering  in databases, \ the players are tuples in the database  ${D}$. We also have a
Boolean query \ $\mc{Q}$, which becomes a   game function, as follows:  \ For \ ${S \subseteq D}$,
\begin{equation*}{\mc{Q}(S) = \left\{\begin{array}{cc} 1 & \mbox{ if } \ S \models \mc{Q}\\
0 & \mbox{ if } \ S \not \models \mc{Q}\end{array}\right.}
\end{equation*}
With this game elements we can define a specific Shapley-value for a database tuple $\tau$:
\begin{equation*}
\nit{Shapley}(D,{\mc{Q}},{\tau}):= \sum_{S\subseteq
  D \setminus \{{\tau}\}} \frac{|S|! (|D|-|S|-1)!}{|D|!}
(\mc{Q}(S\cup \{{\tau}\})-\mc{Q}(S)).
\end{equation*}
If the query is {\em monotone}, i.e. its set of answers never shrinks when new tuples are added to the database, which is the case of conjunctive queries (CQs), among others, the difference $\mc{Q}(S\cup \{{\tau}\})-\mc{Q}(S)$ is always $1$ or $0$, and the average in the definition
of the Shapley-value returns a value between $0$ and $1$. \
This value quantifies the contribution of tuple \ ${\tau}$ \ to the query result. It was introduced and investigated in \cite{LBKS20},
for BCQs and some aggregate queries defined over CQs. We report on some of the findings in the rest of this section. The analysis has been extended to queries with  negated atoms in CQs \cite{ester2}.

A main result obtained in \cite{LBKS20} is in relation to the complexity of computing this Shapley score. It is the following
{\em Dichotomy Theorem}:  \ For ${\mc{Q}}$ a BCQ without self-joins, if \ ${\mc{Q}}$ \ is {\em hierarchical}, then ${\nit{Shapley}(D,\mc{Q},\tau)}$ can be computed in polynomial-time
(in the size of $D$); otherwise, the problem is \ {$\nit{FP}^{\#P}$-complete}. \

Here,  ${\mc{Q}}$ \ is {hierarchical} if for every two existential variables ${x}$ and ${y}$, it holds: \ (a)
{$\nit{Atoms}(x) \subseteq \nit{Atoms}(y)$}, \ or
 {$\nit{Atoms}(y) \subseteq \nit{Atoms}(x)$}, \ or
 {$\nit{Atoms}(x) \cap \nit{Atoms}(y) = \emptyset$}.
 \ For example,  ${\mc{Q}: \ \exists x \exists y \exists z(R(x,y) \wedge S(x,z))}$, for which \
{$\nit{Atoms}(x)$ $ = \{R(x,y),$ $ \ S(x,z)\}$,  \ $\nit{Atoms}(y) = \{R(x,y)\}$,  \ $\nit{Atoms}(z) = \{S(x,z)\}$}, is
hierarchical. \
However, \ ${\mc{Q}^{\nit{nh}}: \ \exists x \exists y({R(x) \wedge S(x,y) \wedge T(y)})}$, for which \
  {$\nit{Atoms}(x) = \{R(x), \ S(x,y)\}$,  \ $\nit{Atoms}(y) = \{S(x,y), T(y)\}$}, is not hierarchical.

These are the same criteria for (in)tractability that apply to BCQs  over probabilistic databases \cite{probDBs}. However, the same proofs do not (seem to) apply.
The intractability result uses query  ${\mc{Q}^{\nit{nh}}}$ \ above, and a
reduction from {counting independent sets in a bipartite graph}.

The {dichotomy results can be extended  to summation} over CQs, with the same conditions and cases. This is because the
Shapley-value, as an expectation, is linear. \
{Hardness extends to aggregates {\sf max}, {\sf min}, and {\sf avg} over non-hierarchical queries}.

For the hard cases, there is an {\em Approximation Result:} \ For every fixed BCQ $\mc{Q}$ (or summation over a  CQ), there is a multiplicative fully-polynomial randomized approximation scheme (FPRAS), $A$, with
$${P(\tau \in D ~|~ \frac{\nit{Shapley}(D,\mc{Q},\tau)}{1+\epsilon} \leq A(\tau,\epsilon,\delta) \leq (1 + \epsilon)\nit{Shapley}(D,\mc{Q},\tau)\}) \geq 1 - \delta}.$$

 A related and popular score, in coalition games and other areas, is the  {\em Bahnzhaf Power Index}, which is similar to the Shapley-value, but the order of players is ignored, by considering subsets of players rather than permutations thereof:
\begin{equation*}{\nit{Banzhaf}(D,\mc{Q},\tau) := \frac{1}{2^{|D|-1}} \cdot \sum_{S \subseteq (D\setminus \{\tau\})} (\mc{Q}(S \cup \{\tau\}) - \mc{Q}(S))}.
\end{equation*}
The Bahnzhaf-index is  also difficult to compute; provably \#P-hard in general. The results in \cite{LBKS20} carry over to this index when applied to query answering in databases.

In \cite{LBKS20} it was proved that the causal-effect score of Section \ref{sec:CE} coincides with the Banzhaf-index, which gives to the former a more fundamental or historical justification.

\section{Score-Based Explanations for Classification}\label{sec:cla}

Let us consider a classifier, $\mc{C}$, that receives a representation of a entity, $\e$, as a record of feature values, and outputs a label, $L(\e)$, corresponding to the possible decision alternatives.
We could see $\mc{C}$ as a black-box, in the sense that only by direct interaction with it,  we have access to its input/output relation. We may not have access to the mathematical classification model inside $\mc{C}$.

\vspace{2mm}
\centerline{\includegraphics[width=5cm]{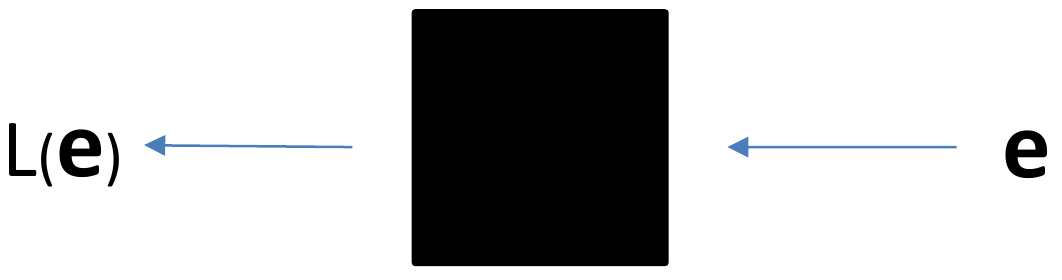}}

\vspace{1mm}
To simplify the presentation we will assume that the entities and the classifier are binary, that is, in the representation ${\mathbf{e}=\langle x_1, \ldots, x_n\rangle}$ of an  entity, the feature values are binary ($0$ or $1$), corresponding to propositional features (false or true, resp.). The label is always $0$ or $1$. \ For example, we could have a client of a financial institution requesting a loan, but the classifier, on the basis of his/her feature values, assigns the label $1$, for rejection. An explanation is requested by the client. Of course, the same situation may occur if we have an explicit classification model, e.g. a classification tree or \ a {logistic regression model}, in which cases, we might be in a better position to given an explanation, because we can inspect the internals of the model \cite{rudin}. However, we will put ourselves in the ``worst scenario" in which we do not have access to the internal model.

An approach to explanations that has become popular, specially in the absence of the model, assigns numerical {\em scores}, trying to answer the question about which   of the feature  values $x_i$ of $\e$ contribute the most to the received label.

Score-based methodologies are sometimes based on counterfactual interventions: What would happen with the label if we change this value, leaving the others fixed? Or the other way around: What if we leave this value fixed, and change the others? The resulting labels can be aggregated, leading to a score for the feature value under inspection.

In the next two sections we briefly introduce two scores. Both can be applied with open-box or black-box models.

\subsection{The $\shap$-score}\label{sec:shap}

 We will consider until further announcement the {\em uniform probability
  space}. Actually, since we consider only binary feature values taking values $0$ or $1$, this is the uniform distribution on  $E = \{0,1\}^n$, assigning probability $P^u(\e) = \frac{1}{2^n}$ to $\e \in E$. One could consider appealing to other, different distributions.

In  the context of classification, the Shapley-value has taken the form of the $\shap$-score \cite{LetA20}, which we briefly introduce. Given the binary classifier, $\mc{C}$, on binary entities, it becomes crucial to identify a suitable game function. In this case, it will be expressed in terms of expected values (not unlike the causal-effect score), which requires an underlying probability space on the population of entities. For the latter we use, as just said, the {\em uniform distribution} over $\{0,1\}^n$.

Given a set of features $\mc{F} = \{F_1, \ldots, F_n\}$, and an entity ${\mathbf{e}}$ whose label is to be explained,  the set of players $D$ in the game is $\mc{F}(\e) :=  \{F(\mathbf{e})~|~  F \in \mc{F}\}$, i.e. the set of feature values of  $\mathbf{e}$. Equivalently, if $\e = \langle x_1, \ldots, x_n\rangle$, then $x_i = F_i(\e)$. We assume these values have implicit feature identifiers, so that duplicates do not collapse, i.e. $|\mc{F}(\e)| = n$.  The game function is defined as follows. \ For $S \subseteq \mc{F}(\e)$,

\vspace{2mm}
\centerline{${\mc{G}_\mathbf{e}(S) := \mathbb{E}(L(\mathbf{e'})~|~\mathbf{e'}_{\!S} = \mathbf{e}_S)}$,}

\vspace{2mm} \noindent where $\mathbf{e}_S$: is the projection of $\e$ on $S$. That is, the expected value of the label for entities $\e'$ when their feature values  are fixed and equal to those in in $S$ for $\e$. Other than that, the feature values of $\e'$ may independently vary over $\{0,1\}$.

Now, one can instantiate the general expression for the Shapley-value in (\ref{eq:sh}), using this game function, as ${\nit{Shapley}(\mc{F}(\e),\mc{G}_\mathbf{e},F(\mathbf{e}))}$,
obtaining, for a particular feature value $F(\e)$:
\begin{eqnarray*}
\shap(\mc{F}(\e),\mc{G}_\mathbf{e},F(\mathbf{e})) &:=& \sum_{S\subseteq
  \mc{F}(\e) \setminus \{F(\mathbf{e})\}} \frac{|S|! (n-|S|-1)!}{n!} \times \\
&&\hspace{-0.5cm}(\mathbb{E}(L(\mathbf{e}'|\mathbf{e}'_{S\cup \{F(\mathbf{e})\}} = \mathbf{e}_{S\cup \{F(\mathbf{e})\}})\ - \ \mathbb{E}(L(\mathbf{e}')|\mathbf{e}'_S = \mathbf{e}_S)).
\end{eqnarray*}
Here, the label ${L}$ acts as a Bernoulli random variable that takes values through the classifier.
\ We can see that the $\shap$-score is a weighted average
of differences of expected values of the labels \cite{LetA20}.

\subsection{The ${\sf RESP}$-score}\label{sec:resp}

In the same setting of Section \ref{sec:shap}, let us  consider the following score introduced in \cite{BLSSV20}. \ For $F \in \mc{F}$, and an entity $\e$ for which we have obtained label $1$, the ``negative" outcome one would like to see explained:
\begin{equation}
\mbox{\sf \small COUNTER}(\mathbf{e},F) := L(\mathbf{e}) - \mathbb{E}(L(\mathbf{e'})~|~\mathbf{e'}_{\!\!_{\mc{F}\smallsetminus\{F\}}} = \mathbf{e}_{_{\mc{F}\smallsetminus\{F\}}}). \label{eq:count}
\end{equation}
This score measures the expected difference between the label for $\e$ and those for entities that coincide in feature values everywhere with $\e$ but on feature $F$. Notice the essential counterfactual nature of this score, which is reflected in all the possible hypothetical changes of features values in $\e$.

The {\mbox{\sf \small COUNTER}-score can be applied in same scenarios as $\shap$, it is {easier  to compute}, and gives {reasonable and intuitive results}, and also behaves well in experimental comparisons with other scores \cite{BLSSV20}.
As with the $\shap$-score, one could consider different underlying probability distributions (c.f. \cite{BLSSV20} for a discussion).
Again, so as for \shap, there is no need to access the internals of the classification model.

 One problem with {\mbox{\sf \small COUNTER} is that changing a single value, no matter how, may not switch the original label, in which case
no explanations are obtained. In order to address this problem, we can bring in {\em contingency sets} of feature values, which leads to the
${\mbox{\sf \footnotesize RESP}}$-score} introduced in \cite{BLSSV20}. We just give the idea and a simplified version of it by means of an  example.

 \begin{example} In the picture below, the black box is the classifier. Entities have three feature values. The table on the right-hand side shows all the possible entities with their labels.
\ We want to explain the label $1$ obtained by entity $\e_1$.

\vspace{-4mm}
\begin{multicols}{2}

\vspace*{-8mm}\hspace*{-1.3cm}\includegraphics[width=10cm]{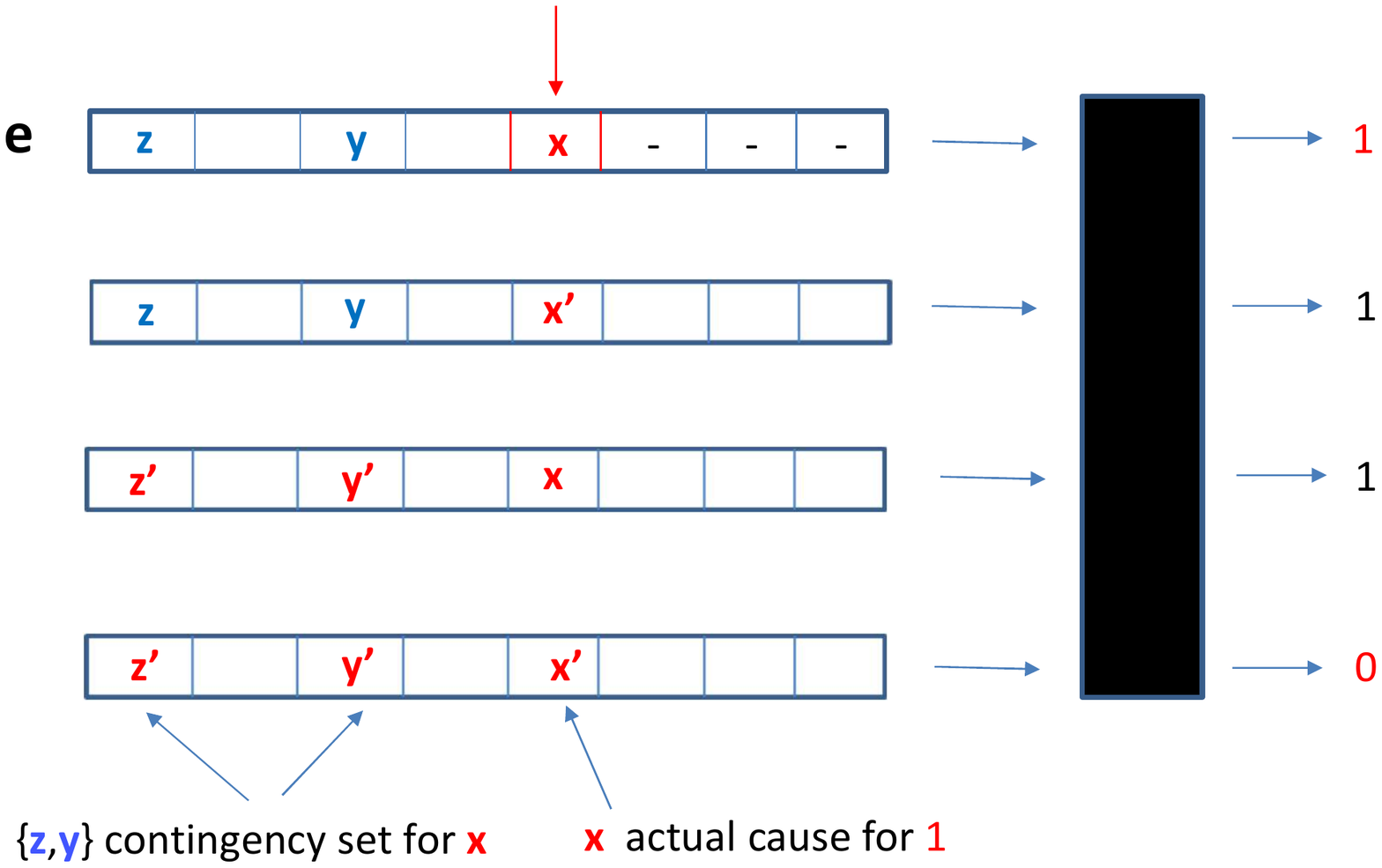}

\hspace*{2.6cm}$\mc{C}$

\vspace{1mm}
\hspace*{1cm}\begin{tabular}{|c|| c|c|c||c|}\hline
entity (id) & ${F_1}$ & ${F_2}$ & $F_3$ & $L$\\ \hline
${\e_1}$ & {0} & {1 }& {1} &{1}\\ \hline
$\e_2$ & 1 & 1 & 1 &1\\
$\e_3$ & 1 & 1 & 0 &1\\
${\e_4}$ & {\bf 1} & {\bf 0} & 1 &{0}\\
$\e_5$ & 1 & 0 & 0 &1\\
$\e_6$ & 0 & 1 & 0 &1\\
${\e_7}$ & 0 & {\bf 0} & 1 &{0}\\
$\e_8$ & 0 & 0 & 0 &0\\ \hline
\end{tabular}
\end{multicols}

\vspace{-2.5cm}Through counterfactual interventions we  change feature values in $\e_1$,    trying  to change the label to $0$. This process is described in the figure above, on the left-hand side, where we are attempting to quantify the contribution of value $\mathbf{ x} = F(\mathbf{e}_1)$. \ Let us assume that by changing $\mbf{ x}$ into any $\mathbf{ x'}$, we keep obtaining label $1$. So, we leave $\mathbf{x}$ as it is, and consider changing
 other original values, $\mathbf{ y}$ and $\mathbf{ z}$, into $\mathbf{y'}$ and $\mathbf{z'}$, still getting $1$. However, if we now, in addition, change $\mathbf{ x}$ into $\mathbf{ x'}$, we get label $0$. \ Then, in the spirit of actual causality, as seen in Section \ref{sec:causal}, we can say that the feature value $\mathbf{ x}$ is an actual cause for the original label $1$, with $\mathbf{ y}$ and $\mathbf{ z}$ forming a contingency set for $\mathbf{ x}$; in this case,  of size $2$.

On this basis, we can define \cite{ruleml20}: \ (a) ${\mathbf{x}}$ \ is a {\em counterfactual explanation} for \ ${L(\mathbf{e}) =1}$ \ if \ ${L(\mathbf{e}\frac{{\mathbf{x}}}{\mathbf{x}^\prime}) = 0}$, \ for some ${\mathbf{x}^\prime \in \nit{Dom}(F)}$ (the domain of feature $F$). \ (b)
 ${\mathbf{x}}$ \ is an {\em actual explanation} for \ ${L(\mathbf{e}) =1}$ \ if there is a set  of values ${\mathbf{Y}}$ in ${\mathbf{e}}$, with ${{\mathbf{x}} \notin \mathbf{Y}}$, and new values ${\mathbf{Y}^\prime \cup \{\mathbf{x}^\prime\}}$, such that \ ${L(\mathbf{e}\frac{\mathbf{Y}}{\mathbf{Y}^\prime}) = 1}$  and \ ${L(\mathbf{e}\frac{{\mathbf{x}}\mathbf{Y}}{\ \mathbf{x}^\prime\mathbf{Y}^\prime}) = 0}$. \ Here, as usual, $\frac{{\mathbf{x}}}{\mathbf{x}^\prime}$,  denotes the replacement of value ${\mathbf{x}}$ by ${\mathbf{x}'}$, and so on.

Contingency sets may come in sizes from $0$ to $n-1$ for feature values in records of length $n$. Accordingly, we can define for the actual cause $\mbf{x}$: If ${\mbf{Y}}$ is a minimum size contingency set for $\mbf{x}$, ${\mbox{\sf \small RESP}(\mbf{x}) := \frac{1}{1 + |\mbf{Y}|}}$; \ and as $0$ when $\mbf{x}$ is not an actual cause. \ This score can be formulated in terms of expected values, generalizing expression (\ref{eq:count}) through the introduction of contingency sets \cite{BLSSV20}.

Coming back to the entities in the figure above, due to \ ${\e_7}$, \ ${F_2(\e_1)}$ \ is  counterfactual explanation; \ with  ${\mbox{\sf \small RESP}({F_2(\e_1)}) = 1}$. \
 Due to \ ${\e_4}$, \ ${F_1(\e_1)}$ \ is  actual explanation; \ with $\{{F_2(\e_1)}\}$ as contingency set, and
\  ${\mbox{\sf \small RESP}({F_1(\e_1)}) = \frac{1}{2}}$. \boxtheorem
\end{example}

\section{Bringing-In Domain Knowledge}\label{sec:cons}

The uniform space gives equal probability to each entity in the underlying population. One can argue that this is not realistic, in that certain combinations of feature values  may be more likely than others; or that certain correlations among them exist. One can consider  assigning or modifying  probabilities in the hope of capturing correlations and logical relationships between feature values.

\subsection{Empirical Distributions}

An alternative consists in using an {\em empirical distribution} as a proxy. In this case we have a sample $S \subseteq E$ \ (we could have repetitions, but we do not consider this case here). The probability of $\e \in E$, is given by:
\begin{equation}
    P_S(\e) := \left\{ \begin{array}{cl}
                                                      \frac{1}{|S|}& \mbox{ if } \e \in S\\
                                                      0& \mbox{ if } \e \notin S\end{array} \right.
\end{equation}
The empirical distribution  was used in \cite{BLSSV20} to compute the $\shap$-score.   More precisely, the entities in $S$ come with labels obtained via the classifier $\mc{C}$; and the score is computed with expectations directly with the entities in $S$, with their labels. The empirical distribution may be better at capturing correlations.

\subsection{Imposing  Constraints}

One can introduce {\em constraints} that prohibit certain combinations of values, in the spirit of {\em denial constraints} in databases, but in this case admitting positive and negative atoms. For example, we may not want the combination of {\em ``The age is not greater than 20"} and {\em ``Gets an account overdraft above \$50M"} to hold simultaneously.

    These constraints, which are satisfied or violated by a single entity at a time,  are of the form:
\begin{equation}
\chi: \ \neg (\bigwedge_{i \in S} F_i \wedge \bigwedge_{j \in S'} \bar{F}_j), \label{eq:cons}
\end{equation}
where $S \cup S' \subseteq F$, \ $S \cap S' = \emptyset$, and $F_i, \bar{F}_j$ mean that features $F_i, F_j$ take values $1$ and $0$, resp.  In the example, it would be of the form $\neg (\overline{\nit{Age}} \wedge \nit{OverDr50M})$.  The events, i.e. subsets of $E$, associated to the violation of $\chi$ should get zero probability.

A way to accommodate a constraint, $\chi$, consists in defining an event associated to it:
\begin{equation*}
A(\chi) = \{\e \in E~|~ \e \models \chi\},
\end{equation*}
where $\e \models \chi$ has the obvious meaning of satisfaction of $\chi$ by entity $\e$.

Given the uniform probability space $\langle E, P^u\rangle$, we can redefine the probability in order to enforce $\chi$. For $A \subseteq E$,
\begin{equation}P^u_\chi(A) := P^u(A|A(\chi)) = \frac{P^u(A \cap A(\chi))}{P^u(A(\chi))}.
\end{equation}
Since $\chi$ is logically consistent (it is satisfied by some entities in $E$), the conditional distribution is well-defined.
Notice that the probability of $\chi$'s violation set, i.e. of
$E \smallsetminus A(\chi)$, is now:
$$P^u_\chi(E \smallsetminus A(\chi)) = \frac{P^u(\emptyset)}{P^u(A(\chi))} = 0.$$
This definition can be extended to finite sets, $\Theta$, of constraints, as long as it is consistent (i.e. satisfiable in $E$), by using $\wedge\Theta$, the conjunction of the constraints in $\Theta$: \
    $P^u_{\Theta}(A) := P^u_{\wedge\Theta}(A)$.

Of course, one could go beyond constraints of the form (\ref{eq:cons}), applying the same ideas, and consider any propositional formula that is intended to be evaluated on a single entity at a time, as opposed to considering  combinations of feature values for   different entities.

The resulting modified distributions that accommodate constraints could be used in the computation of any of the scores expressed in terms of expected values (or in probabilistic terms, in general).

\ignore{
\section{Discussion}\label{sec:disc}

In this work we have considered the uniform probability space. There are other candidates, such as the ``product space" determined by the product of the marginal probabilities of the involved features. In \cite{deem20} it is established that the computation of the $\shap$-score is $\#P$-complete on this space, when the classifier is given by a Monotone-2CNF propositional circuit. \ Since the computational results on the computation of the $\shap$-score for the uniform space may not carry over to the product space, there is lot of open space for future research.

In Section f{sec:cons}, we introduced constraints. They are hard constraints in the sense that they have to be satisfied, which is reflected in the modified probability distribution. It would be interesting to consider {\em soft constraints}, which correspond to making certain events, i.e. subsets of $E$, take a probability above or below a certain threshold. The probability distribution on $E$ would have to be modified accordingly.
}

\section{Final Remarks}\label{sec:last}

 Explainable AI  (XAI)  is an effervescent area of research.
  Its relevance can only grow considering that legislation around explainability, transparency and fairness of AI/ML systems is being produced and enforced.
There are different approaches and methodologies in relation to explanations, and causality, counterfactuals and scores have a relevant role to play.

 Much research is still needed on the {use of contextual, semantic and domain knowledge}. Some approaches may be more appropriate, e.g. declarative ones \cite{ruleml20}.

 Still fundamental research is needed on {\em what is a good explanation}, and in particular, on
what are the {desired properties of an explanation score}. After all, the original, general
Shapley-value emerged from a list of {\em desiderata} in relation to coalition games. Although the Shapley value is being used in XAI, in particular in its \shap \ incarnation, there could be a different and specific  set of desired properties of explanation scores that could lead to a still undiscovered explanation-score.

 \vspace{3mm} \noindent {\bf Acknowledgments: } \ L. Bertossi is a member of the Academic Network of RelationalAI Inc.,
 where his interest in explanations in ML started.


\begin{thebibliography}{10}

\ignore{\bibitem{ACMS20}
A.~Amarilli, F.~Capelli, M.~Monet, and P.~Senellart.
\newblock Connecting knowledge compilation classes and width parameters.
\newblock {\em Theory Comput. Syst.}, 64(5):861--914, 2020.}

\bibitem{bertossiSynth}
Bertossi.~L. \ {\em Database Repairing and Consistent Query Answering}. \ Synthesis Lectures in Data Management. Morgan \& Claypool, 2011.

\bibitem{tocs}
Bertossi,~L. and Salimi,~B. From causes for database queries to repairs and model-based diagnosis and back. {\em Theory of Computing Systems}, 2017,  61(1):191-232.

\bibitem{flairsExt}
Bertossi,~L. and Salimi,~B. \ Causes for query answers from databases: datalog abduction, view-updates, and integrity constraints. \ {\em Int. J. Approximate Reasoning}, 2017, 90:226-252.

\bibitem{foiks18}
Bertossi,~L. \ Characterizing and computing causes for query answers in databases from database repairs and repair programs. Proc. FoIKs, 2018, Springer LNCS 10833, pp. 55-76. Extended version posted as  arXiv:1712.01001, 2020.

\bibitem{BLSSV20}
Bertossi,~L., Li,~J., Schleich,~M., Suciu,~D. and Vagena,~Z. \
\newblock Causality-based explanation of classification outcomes.
\newblock In {\em Proceedings of the Fourth Workshop on Data Management for
  End-To-End Machine Learning, DEEM@SIGMOD 2020}, pages 6:1--6:10, 2020.

  \bibitem{ruleml20}
Bertossi,~L. \ An ASP-based approach to counterfactual explanations for classification. \ To appear in Proc. RuleML-RR'20. Arxiv:2004.13237, 2020.

\bibitem{BetA84}
L.~Breiman, J.~Friedman, C.~J. Stone, and R.~A. Olshen.
\newblock {\em Classification and regression trees}.
\newblock CRC press, 1984.

\bibitem{lineage}
Buneman,~P.,  Khanna,~S.  and  Tan,~W.~C. \   Why  and  where:  a characterization of data provenance. \ Proc. ICDT, 2001, pp. 316-330.

\bibitem{CetA18}
Chen,~C., Lin,~K., Rudin,~C., Shaposhnik,~Y., Wang,~S. and Wang,~T. \
\newblock An interpretable model with globally consistent explanations for
  credit risk.
\newblock {\em CoRR}, abs/1811.12615, 2018.

\bibitem{CH04}
Chockler,~H. and Halpern,~J. \
\newblock Responsibility and blame: A structural-model approach.
\newblock {\em J. Artif. Intell. Res.}, 22:93--115, 2004.

\ignore{\bibitem{D01}
A.~Darwiche.
\newblock On the tractable counting of theory models and its application to
  truth maintenance and belief revision.
\newblock {\em J. Appl. Non Class. Logics}, 11(1-2):11--34, 2001.

\bibitem{DBLP:journals/jair/DarwicheM02}
A.~Darwiche and P.~Marquis.
\newblock A knowledge compilation map.
\newblock {\em J. Artif. Intell. Res.}, 17:229--264, 2002.  }

\bibitem{DBLP:journals/mor/DengP94}
Deng,~X. and Papadimitriou,~C. \
\newblock On the complexity of cooperative solution concepts.
\newblock {\em Math. Oper. Res.}, 19(2):257--266, 1994.

\bibitem{FK92}
Faigle,~U. and Kern,~W. \
\newblock The shapley value for cooperative games under precedence constraints.
\newblock {\em International Journal of Game Theory}, 21:249--266, 1992.

\bibitem{HP05}
Halpern,~J. and Pearl,~J. \
\newblock Causes and explanations: A structural-model approach. part i: Causes.
\newblock {\em The British journal for the philosophy of science},
  56(4):843--887, 2005.

\bibitem{HK10}
Hunter,~A. and Konieczny,~S. \
\newblock On the measure of conflicts: Shapley inconsistency values.
\newblock {\em Artif. Intell.}, 174(14):1007--1026, 2010.

\ignore{
\bibitem{DBLP:journals/mst/JhaS13}
A.~K. Jha and D.~Suciu.
\newblock Knowledge compilation meets database theory: Compiling queries to
  decision diagrams.
\newblock {\em Theory Comput. Syst.}, 52(3):403--440, 2013. }

\bibitem{LBKS20}
Livshits,~E., Bertossi,~L., Kimelfeld,~B. and Sebag,~M. \
\newblock The Shapley value of tuples in query answering.
\newblock In {\em 23rd International Conference on Database Theory, {ICDT}
  2020, March 30-April 2, 2020, Copenhagen, Denmark}, volume 155, pages
  20:1--20:19, 2020.

\bibitem{LHdR19}
Lucic,~A.,  Haned,~H.  and de~Rijke,~M. \
\newblock Explaining predictions from tree-based boosting ensembles.
\newblock {\em CoRR}, abs/1907.02582, 2019.

\bibitem{LetA20}
Lundberg,~S., Erion,~G., Chen,~H., DeGrave,~A., Prutkin,~J.,  Nair,~B., Katz,~R.,
  Himmelfarb,~J., Bansal,~N.  and  Lee,~S.-I. \
\newblock From local explanations to global understanding with explainable ai
  for trees.
\newblock {\em Nature machine intelligence}, 2(1):2522--5839, 2020.

\bibitem{LL17}
Lundberg,~S. and Lee,~S. \
\newblock A unified approach to interpreting model predictions.
\newblock In {\em Advances in Neural Information Processing Systems 30: Annual
  Conference on Neural Information Processing Systems 2017, 4-9 December 2017,
  Long Beach, CA, {USA}}, pages 4765--4774, 2017.



  \bibitem{suciu}
Meliou,~A., Gatterbauer,~W.,
 Moore,~K.~F. and  Suciu,~D.
\newblock The complexity of causality and responsibility for query answers and
  non-answers.
\newblock Proc. VLDB, 2010, pp. 34-41.

\bibitem{suciuDEBull}
Meliou,~A., Gatterbauer,~W., Halpern,~J.Y., Koch,~C., Moore,~K.~F. and  Suciu,~D. \
Causality in databases. \ {\em IEEE Data Eng. Bull.}, 2010, 33(3):59-67.



\bibitem{M97}
Mitchell,~T.~M. \
\newblock {\em Machine Learning}.
\newblock McGraw Hill series in computer science. McGraw-Hill, 1997.

\bibitem{molnar}
Molnar,~C. \ {\em Interpretable Machine Learning:
A Guide for Making Black Box Models Explainable}. \ {\text https://christophm.github.io/interpretable-ml-book}, 2020. 


\bibitem{DBLP:books/cu/NRTV2007}
Nisan,~N., Roughgarden,~T., Tardos,~E. and  Vazirani,~V.~V. (eds.)
\newblock {\em Algorithmic Game Theory}.
\newblock Cambridge University Press, 2007.

\bibitem{ester2}
Reshef,~A., Kimelfeld,~B. and Livshits,~E. \
The impact of negation on the complexity of the shapley value in conjunctive queries. Proc. PODS 2020, pp. 285-297.

\bibitem{rudin}
Rudin,~C. \ Stop explaining black box machine learning models for high stakes decisions and use interpretable models instead. \ {\em Nature Machine Intelligence}, 2019,
1:206-215. \ Also arXiv:1811.10154,2018.

\bibitem{R88}
Roth,~A.~E. (ed.) \
\newblock {\em The Shapley Value: Essays in Honor of Lloyd S. Shapley}.
\newblock Cambridge University Press, 1988.

\bibitem{tapp16}
Salimi, B., Bertossi, L., Suciu, D. and Van den Broeck, G. \ Quantifying causal effects on query answering in databases. Proc. 8th USENIX Workshop on the Theory and Practice of Provenance (TaPP), 2016.

\bibitem{S53}
Shapley,~L.~S. \
\newblock A value for n-person games.
\newblock {\em Contributions to the Theory of Games}, 2(28):307--317, 1953.

\bibitem{probDBs}
Suciu, D., Olteanu, D., Re, C. and Koch, C. \ {\em Probabilistic Databases}.
Synthesis Lectures on Data Management, Morgan \& Claypool, 2011.

\end{thebibliography}
\end{document}